# Co-contributorship Network and Division of Labor in Individual Scientific Collaborations

Chao Lu[1], Yingyi Zhang[2], Yong-Yeol Ahn[3], Ying Ding[4], Chenwei Zhang[3, 5], Dandan Ma[6, *]

*1. Business School, Hohai University, Nanjing, Jiangsu, China*

*2. School of Economics and Management, Nanjing University of Science and Technology, Nanjing, Jiangsu, China*

*3. School of Informatics, Computing, and Engineering, Indiana University, Bloomington, Indiana, USA*

*4 School of Information, The University of Texas at Austin, Austin, Texas, U.S.A.*

*5. Qiqihar Institute of Engineering, Qiqihar, Heilongjiang, China*

*6. School of Information Engineering, Nanjing University of Finance and Economics, Nanjing, Jiangsu, China*

**Correspondence concerning this article should be addressed to Dr. Ma Dandan**, Email: jane.ma1030@gmail.com

# Co-contributorship Networks and Division of Labor in Scientific Collaboration

**Abstract**: Collaborations are pervasive in current science. Collaborations have been studied and encouraged in many disciplines. However, little is known how a team really functions from the detailed division of labor within. In this research, we investigate the patterns of scientific collaboration and division of labor within individual scholarly articles by analyzing their co-contributorship networks. Co-contributorship networks are constructed by performing the one-mode projection of the author-task bipartite networks obtained from 138,787 papers published in *PLoS* journals. Given a paper, we define three types of contributors: Specialists, Team-players, and Versatiles. Specialists are those who contribute to all their tasks alone; team-players are those who contribute to every task with other collaborators; and versatiles are those who do both. We find that team-players are the majority and they tend to contribute to the five most common tasks as expected, such as "data analysis" and "performing experiments". The specialists and versatiles are more prevalent than expected by a random-graph null model. Versatiles tend to be senior authors associated with funding and supervisions. Specialists are associated with two contrasting roles: the supervising role as team leaders or marginal and specialized contributions.

## 1. INTRODUCTION

In science, many solitary individuals' efforts are appreciated and emphasized. For example, people often link some individuals' names with great findings, such as Sigmund Freud with the Interpretation of Dreams, Albert Einstein with the Theory of Relativity, and John von Neumann with the Theory of Games and Economic Behavior. However, more scientific and industrial progress that has made history come from powerful collaborations. In recent decades more and more research has been conducted by groups of scholars. For example, thanks to the joint efforts of Watson, Crick, Franklin, and Wilkins, the double-helix structure of DNA was discovered, which is fundamental to the modern biotechnology (Science History Institute, 2017). Later since 1990, twenty institutions from six countries participated in the great exploration of sequence and map of all human genes, known as Human Genome Project. After more than 20 years' hard working, scientists are able to present nature's complete genetic blueprint of a human being; such findings greatly contribute to treat, cure, and prevent various human diseases (National Human Genome Research Institute, 2019). With the increasing complexity of problems to solve, such as designing the new functional protein or developing self-driving cars, collaboration is necessary. When checking up recent leading studies, we will find that many of them have a long list of contributors or acknowledgment, which reveal the intensity of collaborations. Collaboration can bring a lot of advantages, for example, it can decrease the cost (Katz & Martin, 1997), bring in more expertise thus boost up the efficiency (Goffman & Warren, 1980), and increase the scientific popularity, visibility, and recognition (Price & Beaver, 1966; O'Connor, 1970). Collaborations make impossible possible. Many believe the power of scientific collaborations and have spent efforts to find collaborators and work in teams (Fox & Faver, 1984).

The increasing demand for scientific collaboration has attracted numerous scholars to study the mechanism of collaboration from different perspectives, such as bibliometrics (e.g., Ding, Foo, & Chowdhury, 1998; Glänzel, 2002), social network analysis (e.g., Barabâsi et al., 2002; Newman, 2004; Zhang, Bu, Ding, & Xu, 2018), and qualitative approaches (e.g., Birnholtz, 2006; Hara, Solomon, Kim, &

Sonnenwald, 2003; Lee & Bozeman, 2005;). Despite some differences, in bibliometrics, most of the researches use co-authorship to measure scientific collaborations (Milojević, 2010). Studies using co-authorship usually assume that each collaborator shares equal contributions to their scientific works and based on that they build co-author networks to study scientific collaboration (e.g., Birnholtz, 2006; Chompalov, Genuth, & Shrum, 2002; Newman, 2004). However, little is known about how each collaborating individual works in a team. Do they still collaboratively complete the whole procedure of work? Or do they divide the labor thus each only accomplishes certain tasks within a team then the final goal is achieved by assembling all these tasks?

Early in about fourth century BC, Plato stated the importance of the *division of labor* for the emergence of cities in his *Republic*; Xenophon also noticed the existence of specialization and mentioned the *division of labor* enhances productivity in his *Cyropaedia*. Centuries ago, Smith (1776) discovered that *division of labor*, a proper division and combination of different operations in manufacturing, improves the efficiency of production; it further impacts our whole modern society as it shapes how people are interacting with each another to achieve goals (e.g., Durkheim, 1933; Earley, 1993; Ezzamel & Willmott, 1998).For this concern, in a team of various forms, how the tasks are divided and performed among the members determines its performance (Delfgaauw, Dur, & Souverijn, 2018). Thus, it is in a great need to examine the mechanisms of teamwork via investigating the division of labor in teamwork (or collaboration). In scientific collaborations, faster and greater scientific innovations are always encouraged. Therefore, it is of greater value to know how to achieve a successful scientific collaboration by proper division of labor—which tasks each team member should take on and what kind of collaborations enables collaborators to better achieve their scientific goals (Hara et al., 2003; Ilgen et al., 2005; Leahey & Reikowsky, 2008; Melin, 2000).

Currently, only a few have examined the scientific collaboration at the level of task assignments (Larivière et al., 2016; Jabbehdari & Walsh, 2017; Corrêa Jr, Silva, Costa, & Amancio, 2017; Yang, Wolfram, & Wang, 2017). However, their focus is on tasks globally, rather than from the perspective of interactions within each team. For example, Jabbehdari & Walsh (2017) estimated the likelihood of specialist authors by checking the authors' tasks via a survey on 8,864 papers. Yang *et al.* (2017) analyzed the relationship between authors' tasks in the contribution lists and their positions in the bylines. There is a lack of research investigating the detailed division of labor within every single collaboration. Here, we comprehensively analyze how members in a team divide the labor by recognizing and examining different roles they conduct on a large-scale dataset. Our study helps understand scientific collaborations in depth by revealing the fundamental mechanisms of how collaborative teams function. Inspired by the approaches to using the contribution statements to study author contribution patterns (Corrêa *et al.*, 2017; Larivière et al., 2016; Sauermann & Haeussler, 2017), we analyze the networks of authors and tasks in more than 130,000 papers published in *PLoS* journals. First, we study the density of the co-contributorship networks, which reflects the degree of labor division. We then define three types of author contributions—*specialists*, *team-players*, and *versatiles*—based on the co-contributorship networks, and examine the abundance of these types of contributors. We find that team-players tend to contribute to the five most common tasks, such as "data analysis" and "performing experiments". Versatiles tend to be senior authors associated with funding and supervisions. Specialists are associated with two contrasting roles: the supervising role as a team leader or marginal and specialized contribution. These features will also facilitate us further assess the division of labor and specialization in teams in the future.

# 2. RELATED WORK

## 2.1. Division of Labor in Teamwork Studies

Teamwork is a complex process that involves interactions between members with different expertise and skills with a spectrum of degrees of division of labor. Ilgen *et al.* (2005) argued that the classic IPO (input-process-output) model is insufficient to describe the process of teamwork, due to its complexity. LePine *et al.* (2008)'s meta-analysis found that teamwork generally includes three general processes: mission analysis, action process, and interpersonal process; each of them includes several sub-processes. In the action process, Earley (1993) observed that the psychological statues of team members can affect their diverse collaboration patterns with others, individually or collectively. Studies also extend to classifying task types (e.g., Salas, Sims, & Burke (2005) and team roles (e.g., Belbin, 2012). For instance, Belbin found that a team full of "Apollos" (i.e., geniuses) usually exhibits terrible performance, and that role allocation is necessary for successful teamwork.

Smith (1776) argued that division of labor is a strong impetus for increased productivity and specializations. For example, factory workers can be distributed to specific tasks in the pipeline, so that they can be more concentrated on fine-grained tasks and improve their skills (Leroy, 2009). The degree of division of labor was believed to be limited only by the number of laborers in the market (Stigler, 1951). Meanwhile, if the tasks are complex and interdependent, the coordination cost can be a significant limiting factor on specialization (Becker & Murphy 1992). Therefore, the extent of division of labor may be largely affected by the nature of the tasks.

## 2.2. Scientific Collaboration and Division of Labor

Scientific collaboration as a particular form of teamwork mainly focusing on scientific activities with high intelligence and innovation increasingly prevails in academia (Fox, & Faver, 1984; Guimerà, Uzzi, Spiro, & Amaral, 2005; Katz, & Martin, 1997; Larivière et al., 2015; Wuchty, Jones, & Uzzi, 2007). In this form of teamwork, division of labor is commonly suggested by some studies (Birnholtz, 2006; Fox, & Faver, 1984; Kraut, Galegher, & Egido, 1987; Leahey & Reikowsky, 2008). For example, Merlin (2000) classified scientific teams into two categories: one where everyone in the team is given a clear task assignment and the other where everyone works together. The two types are defined in a similar way by Hara (2003) as "complementary" and "integrative" teams. Chompalov *et al.* (2002) classified teams into four categories based on their topological features: bureaucratic, leaderless, non-specialized, and participatory teams.

Current studies investigate scientific collaboration via co-authorship network analysis (Ahuja, 2000; Yan & Ding, 2009; Newman, 2004; Xie *et al.*, 2018) or using case studies (Amabile et al., 2001) and interviews (Birnholtz, 2006; Chompalov *et al.*, 2002; Chung, Kwon, & Lee, 2016; Fox, & Faver, 1984). These studies reveal several important features in collaborations, such as homophily (Zhang et al., 2018), transitivity (Newman, 2014), and preferential attachment (Milojević, 2010). They also suggest that collaboration improves productivity in science (Lee & Bozeman, 2005) and collaborative researches tend to attract more citations (Larivière et al., 2015). However, such co-authorship studies usually overlooked the division of labor in scientific collaboration at large; and some of them only relied on a limited number of cases. Only a few studies started investigating the tasks conducted by the members of a team. But there is still a lack of research investigating the roles scientists have taken within every single collaboration. These drive us to use author contribution statements embedded in the full text of scientific articles provided by authors to investigate how scientific teams design their tasks and distribute them to collaborators, which is the process of division of labor. So that we can investigate the scientific collaboration between co-authors from the task level and reveal different roles taken by these authors.

## *2.3 Contribution Statement for Scientific Collaboration Studies*

Although the author contribution patterns in scholarly articles have been of interest in scientometrics (e.g., Giles & Councill, 2004; Laudel, 2002), it was the wide adoption of the contribution disclosure policies that enabled large-scale data-driven studies (Allen et al., 2014; Brand et al., 2015; Lariviere et al., 2016). For example, Larivière et al. (2016) examined the forms of division of labor across disciplines, the relationship between contribution types (i.e., writing the paper, performing the experiments, conceiving ideas, analyzing data, and contributing tools) and authors seniority, such as academic ages and that between types of tasks and byline positions. They have found that authors contribute to their studies unevenly across disciplines; that most authors are identified to contribute to writings; and that those who write the papers usually design the studies and those providing materials usually do not perform an experiment and vice versa. They also found that senior authors usually do fewer tasks such as conducting experiments than junior ones but do more tasks such as writing papers and contributing tools and materials. First and last authors usually contribute more tasks than middle ones to their studies. Corrêa *et al.* (2017) placed more emphasis on the relationship between authors' rank positions and the corresponding contributions. They collected author contribution statements in *PLoS ONE*, identified five common tasks and built a bipartite graph for each paper, where authors and the five tasks are the two groups of nodes and an edge between author and task means the author performed the task, treating tasks as equal contributions. Using the average number of tasks authors performed across papers, they found that usually the first and the last authors contribute more to their papers than middle authors, which echoes the findings by Larivière *et al.* (2016). They further identified three general patterns of author contribution with their byline position: the contribution increases with authors' ranks, the contribution decreases with authors' rank, and the contribution decreases then increases with the author's ranks. Sauermann & Haeussler (2017) presented two studies: the first investigated how informative the byline position of an author is about the type and broadness of the author's contribution using more than 12,000 *PLoS ONE* articles; the second reported how author contribution statements are used and scholars' several concerns on authorship and author contribution statement after surveying nearly 6,000 corresponding authors from *PLoS ONE* and *PNAS* (*Proceedings of the National Academy of Sciences of the United States*). The two data sources suggest no significant differences. They also found similar observations that the first and the last authors contribute more than the middle authors to their papers (Corrêa *et al.*, 2017, Larivière *et al.*, 2016). Besides, they also observed that corresponding authors are more likely to be the last authors. First authors usually tend to make more contributions than other authors. When the team gets larger, authors tend to perform fewer tasks, suggesting a stronger degree of division of labor. The top 10% most cited articles maintain similar results from the models generated from the full data set, suggesting the reliability of the author contribution statements from *PLoS ONE* articles.

To sum up, as the division of labor has been an important driving force in the modern society (e.g., Durkheim, 1933; Earley, 1993; Ezzamel & Willmott, 1998), there has been much interest in studying the division of labor or roles in teams, particularly in scientific collaboration teams. The author contribution statements can serve as a good proxy to concretely measure the role allocation and division of labor in scientific collaboration. Given the complex nature of scholarly work, it is of great value to ask how a team can achieve a successful scientific collaboration, how the division of labor occurs in scientific collaboration, and what the patterns of role and labor distribution are (Hara et al., 2003; Ilgen et al., 2005; Leahey & Reikowsky, 2008; Melin, 2000).

# 3. METHODOLOGY

This section explains our dataset and approach to constructing and analyzing the contribution network. Figure 1 illustrates the workflow of this study. First, we collect 138,787 full-text articles from *PLoS*, from which we extract and parse the author contribution statements. From each statement, we extract author-task pairs, which we assemble to construct an author-task contribution network for each paper. The one-mode projection of this network produces a co-contributorship network, from which we define the three types of collaborators. Finally, we further investigate the tasks they partook using content analysis.

*3.1. Data*

**Full-text Data Source**

*PLoS* is one of the largest open-access journal article publishers in the world. Under its authorship policy[1] (which accords with CRediT Taxonomy[2] since 2016), authors are required to state their agreed contributions in their manuscripts. To collect author contribution statements, nearly 170,000 full-text articles published in *PLoS* between 2006 and 2015 have been harvested in the XML format.

In each XML file, the author contribution statement is either embedded in the tag of *"<fn fn-type="con">*" (See the sample article in Figure 1) or the acknowledge part (a few of them).

---

[1] http://journals.plos.org/plosone/s/authorship
[2] http://www.cell.com/pb/assets/raw/shared/guidelines/CRediT-taxonomy.pdf

```
▼<author-notes>
  ▼<corresp id="n101">
     * To whom correspondence should be addressed. E-mail:
     <email xlink:type="simple">eric.spierings@lumc.nl</email>
  </corresp>
  ▼<fn fn-type="con">
    ▼<p>
       Conceived and designed the experiments: EG ES JD. Performed the
       experiments: ES JD MH JP MS. Analyzed the data: EG ES FC JD JP MS.
       Contributed reagents/materials/analysis tools: ES JD MH JP MS. Wrote
       the paper: EG ES.
    </p>
  </fn>
</fn>
```

**Figure 1. An example author contribution statement of a sample article in XML format.**

## Author-task Pairs

We used the XML package in Python 2.7 to extract the contribution statements. Then we used regular expressions to extract author-task pairs from each statement of every paper. Table 1 shows the parsed data from the sample article. We did not separate commonly grouped tasks into sub-tasks. For instance, we consider "Contributed reagents/materials/tools" or "conceived and designed the experiments" as single tasks on their own.

**Table 1. Author-task Pairs of a Sample Article.**

| Id | Authors | Task |
|----|---------|------|
| 1 | EG; ES; JD | Conceived and designed the experiments |
| 2 | ES; JD; MH; JP; MS | Performed the experiments |
| 3 | EG; ES; FC; JD; JP; MS | Analyzed the data |
| 4 | ES; JD; MH; JP; MS | Contributed reagents/materials/tools |
| 5 | EG; ES | Wrote the paper |

The final collected 138,787 articles belong to seven journals in *PLoS* as are shown in Table 2. The table suggests that 90% of articles are from *PLoS ONE*, and the rest 10% belong to the other six journals by *PLoS*. The distribution of those articles by year in our data set is presented in Figure 2, suggesting most of the articles are published before the middle of the year of 2016 when CRediT Taxonomy was launched for regulating author contribution statements.

**Table 2. Journal distribution of our collected author contributions**

| Journal | # of articles | Ratio (%) |
|---------|---------------|-----------|
| *PLoS ONE* | 12,422 | 89.5% |
| *PLoS Genetics* | 3,919 | 2.8% |
| *PLoS Pathogens* | 3,445 | 2.5% |
| *PLoS Computational Biology* | 3,067 | 2.2% |
| *PLoS Neglected Tropical Diseases* | 2,783 | 2.0% |
| *PLoS Biology* | 921 | 0.7% |
| *PLoS Medicine* | 432 | 0.3% |

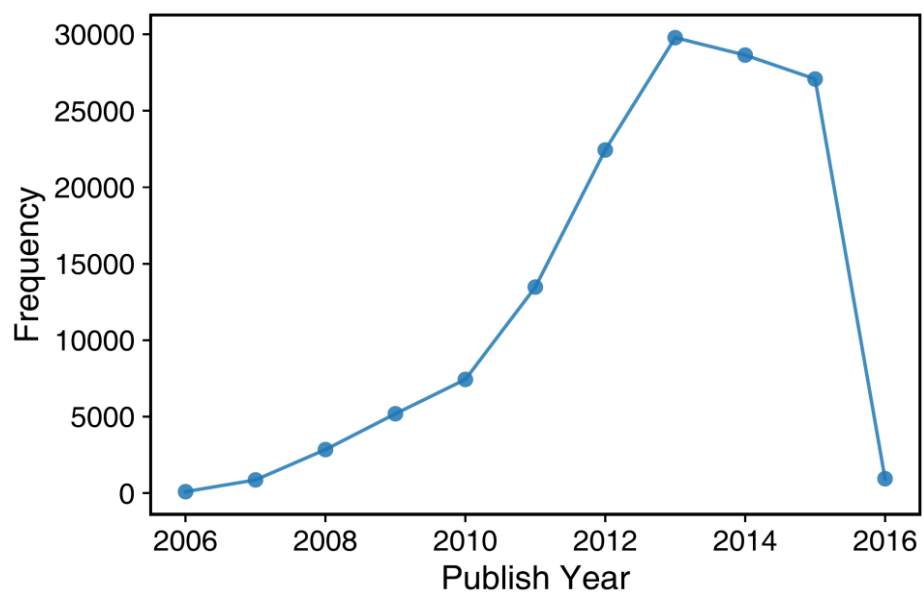

**Figure 2. Yearly distribution of articles in our data set.**

## 3.2. Co-contributorship Network Construction

## <u>Definitions</u>

Figure 2 illustrates different types of collaboration patterns that one can observe from co-contributorship networks. In Figure 3A, every author works collectively on each task, forming a complete graph. Under this scenario, the division of labor does not occur as everyone works on all tasks collectively. By contrast, in Figure 3C, every author works on his/her tasks independently, thus, having a strong division of labor. In our dataset, we expect to see the whole spectrum from no-division to complete division, while most collaborations would occur somewhere in the middle (e.g., Chompalov *et al.*, 2002; Fox, & Faver, 1984; Heffner, 1979).

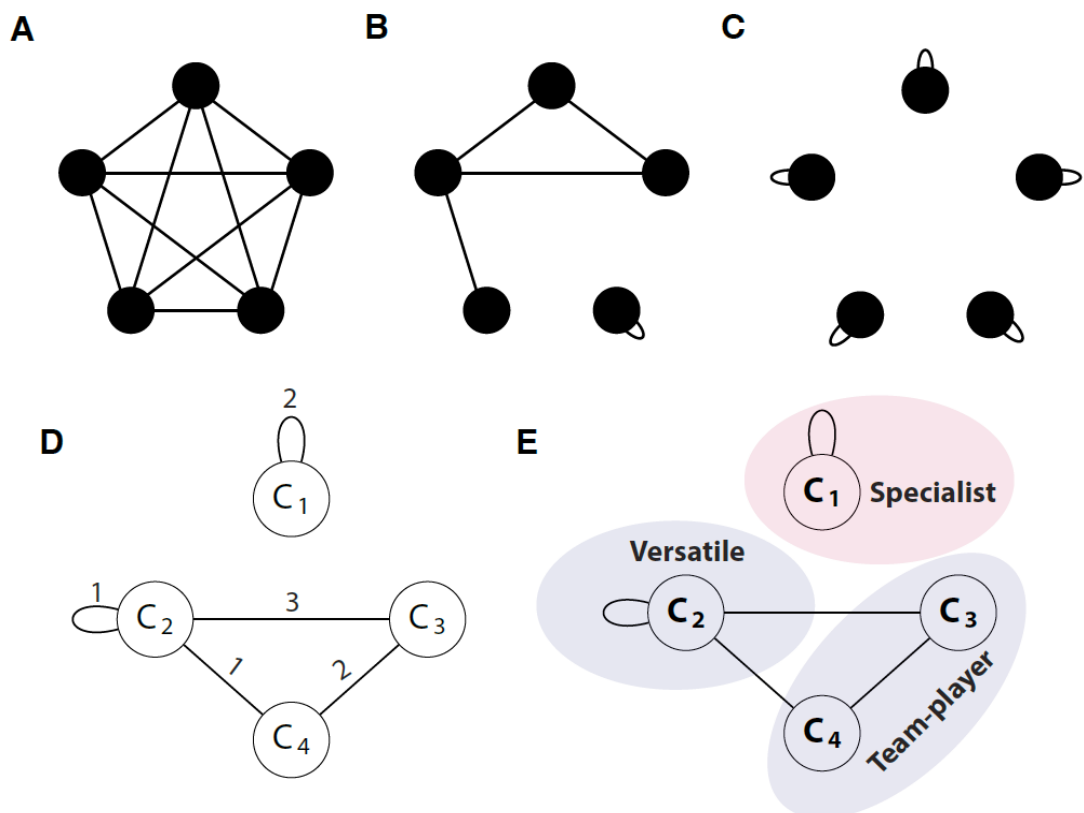

**Figure 3.** Modes of division of labor in teams (A-C); co-contributorship network (D); and types of collaborators (E).

Building on this intuition, we formally consider a weighted undirected co-contributorship network, which can be obtained by performing a special one-mode projection to the author-task network. This process is different from the standard one-mode projection because we also create self-edges if a task is performed by only one person. In the co-contributorship network, each node represents an author (collaborator). An edge between authors means that there is at least one task where two authors collaborated. The weight of each edge represents the number of tasks co-performed by the two authors. If a task is performed by more than two authors, every possible pair of authors will have an edge between them. If a task is performed by a single person, the node (collaborator) will have a self-loop, and its weight is decided by the number of tasks that the author performed independently. As an example, Figure 3D demonstrates a co-contributorship network between four authors in one article. The weight of the edge ($C_2$, $C_3$) is three, which means authors $C_2$ and $C_3$ worked together on three different tasks. The weight for the self-edge of C1 is two, indicating that C1 independently worked on two different tasks alone.

## Types of Collaborators

Based on its connectivity patterns, each node is classified into one of the three roles: team-players, specialists, and versatiles, as shown in Figure 3E. *Team-players* are those who do not have any self-edges; they performed all their tasks with someone else. *Specialists* are those who have only self-edge(s) (e.g., C1); they are those who finish their tasks on their own. *Versatiles* are those who have both self-edges and normal edges (e.g., C2).

## Null Models

To estimate the expected prevalence of each type of authors, we adopt two null models to the author-task contribution networks: the configuration model (Molloy & Reed, 1995) and the Erdős–Rényi random graph model (Erdös & Rényi, 1959). In the configuration model (CFM), the degrees of nodes are fixed while the actual connections are randomized. In creating the networks, we reject the cases with multi-edges. Finally, we project this author-task bipartite graph to a co-contributorship network (see Figure 4-A). For the Erdős–Rényi random graph model (ERM), we fix the number of edges in the author-task graph and randomize the connections without conserving the degree sequences. To make the random graph realistic, we enforce the connectivity of the network—each author node and each task node should have at least one edge. After obtaining an initial random graph, we perform a rejection sampling to obtain a graph where every node has at least one connection (see Figure 4-B). By examining the differences between the actual networks and the two null models, we put our measurements in a reasonable context of random cases.

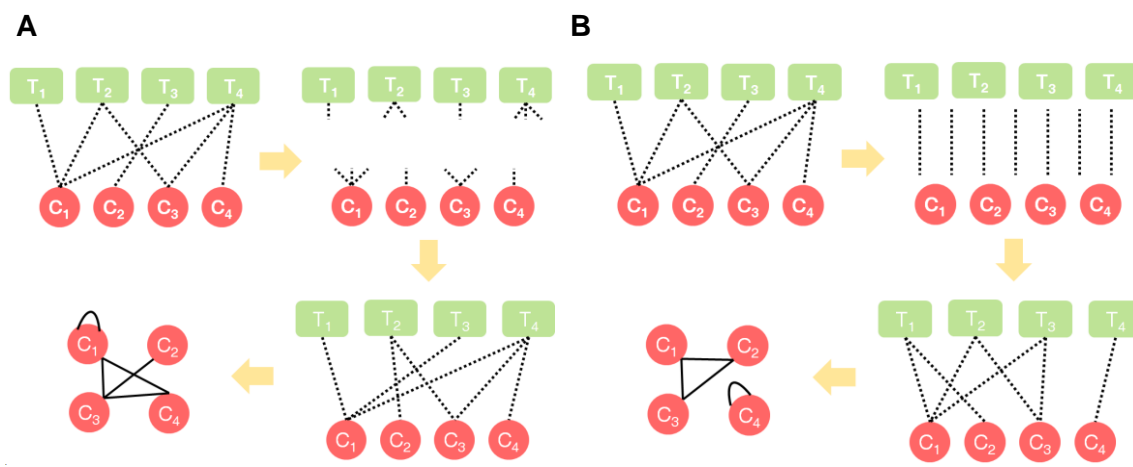

**Figure 4. Producing null models using CFM and ERM. (A) Configuration model (CFM), where the degree sequences on both sides are preserved. (B) Erdős–Rényi random graph model (ERM), where only the total number of edges is preserved with every node has at least one connection.**

3.3 Research Hypotheses

Using the networks we built above, we are to answer three questions concerning the division of labor within teams.

**RQ1**. Is division of labor common in scientific collaborations?

To answer this question, we are to examine the density of each bipartite graph we build compared to the expectation from the ER random graph, maintaining the number of edges between authors and tasks. Ideally, if the division of labor is not necessary for scientific collaboration the graph density distribution of all the networks we build will follow a binomial distribution where the chance to connect an author and a task in an author-task bipartite network is equal. So our first null hypothesis for this question will be:

**H01**. There no difference in the graph density distribution between the real-world author-task bipartite networks and random ones.

**RQ2**. Concerning the three types of collaborators, are they more common than one another in scientific collaboration?

To answer this question, we are to examine the distribution of the three types of authors in their scientific collaborations at a paper level from three perspectives: the existence of the collaborators, the ratios of them in all of the publications, and the ratios of them in the publications with non-teamplayers. we also designed the ER model and the CRF model to remove random factors from the observations. So our null hypotheses for this question will be:

**H2**. The three types of collaborators are equally common in scientific collaborations.

**H3**. The ratios of the three types of collaborators in all publications are equal to each other.

**H4**. The ratios of the three types of collaborators in the publications with non-teamplayers are equal.

**RQ3**. Do the three types of collaborators perform different tasks in their collaborations against each other?

To answer this question, we are to examine the distribution of the tasks that the three types of authors performed in their scientific collaborations at a paper level in two parts: the five common tasks and the rest less frequent tasks in all publications. The ER model and the CRF model will serve to remove random factors from the observations. So our null hypotheses for this question will be:

**H5**. The three types of collaborators contribute equally to the five common tasks in their scientific collaborations.

**H6**. The three types of collaborators contribute equally to the less frequent tasks in their scientific collaborations.

Following the three questions above with 6 null hypotheses, we will use the author co-contributorship networks built from each paper's author contribution statement to address these questions and hypotheses.

# 4. RESULTS AND DISCUSSION

## *4.1. Overview*

More than 90% of the articles in our dataset are written by at least two authors, agreeing with the previous observations that collaborative studies are dominating (Guimerà *et al.*, 2005; Wuchty *et al.*, 2007). 87 percent of articles are written by teams of no more than 10 members; and 99 percent of teams have no more than 20 authors, including 1,370 single-authored articles (8.2%), shown in Figure 5 -A. In the following analyses, we focus on the papers with fewer than 20 authors in our dataset because they occupy the vast majority of the dataset and it is easier to implement the null models for the papers with fewer authors.

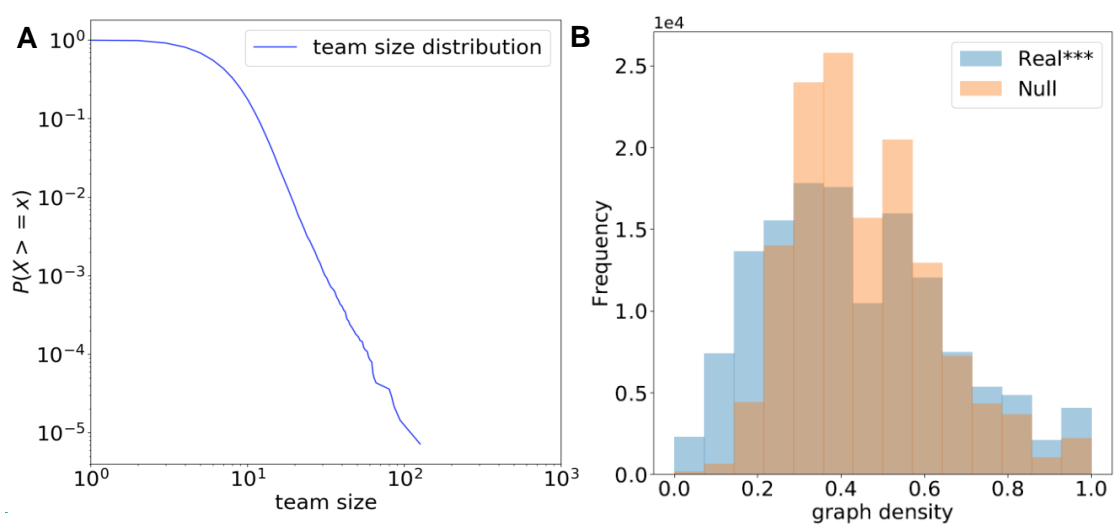

**Figure 5. Author distribution by team size (A); and normalized graph density distribution in author-task bipartite graph (B) (\* on the legend denotes the *p*-value from Kolmogorov–Smirnov test against the null model, \* for *p*-value <0.05, \*\* for *p*-value <0.01, \*\*\* for *p*-value**

**<0.001).**

To depict the division of labor in scientific collaboration, we first calculate the normalized graph density of each author-task bipartite graph compared with one null model generated by Erdős–Rényi random bipartite graph (detailed below). The normalized graph density is calculated using Formula (1):

$$NGD = \frac{k - \max(N_t, N_a)}{N_t \times N_a - \max(N_t, N_a)} \tag{1}$$

Where $k$ represents the number of edges in the graph, $N_t$ number of tasks, and $N_a$ the number of authors. So $N_t \times N_a$ denotes the maximum number of edges in an author-task bipartite graph and $\max(N_t, N_a)$ represents the minimum when all nodes should be connected[3].

To generate our null model here, another Erdős–Rényi random bipartite graph is adopted here, using $G(N_t, N_a, p^j \langle N_a \rangle)$ where $p^j \langle N_a \rangle$ is the probability for an author to perform a task in article $j$ which contains $N_a$ collaborators (Batagelj & Brandes, 2005), estimated by using Formula (2):

$$p^j \langle N_a \rangle = \frac{N_e^j / N_a}{\bar{N}_t} \tag{2}$$

In Formula (2), $N_e^j$ is the number of edges in the author-task bipartite graph of article $j$, $\bar{N}_t$ is the mean number of tasks in all articles with $N_a$ collaborators; and $N_e^j / N_a$ average number of tasks per collaborator performed in $N_a$-author article $j$. Then we use Formula (1) to calculate the normalized graph density for these random graphs.

Figure 5-B shows that the author-task bipartite graphs in our dataset present larger

---

[3] When using Formula (1) to calculate the normalized graph density, if either m or n equals to one, then $m \times n = \max(m, n)$. Under this situation we decide the density is one.

variance in the degree of labor division, compared with a null model that assumes a homogeneous contribution from authors. By examining two ends of the x-axis, it is found that both a strong division of labor and no division of labor are more probable than expected by the homogeneous null model. It might suggest that scientific teams tend to employ a wider variety of collaboration strategy, although our results may be explained by the heterogeneous author degree distribution (i.e. large variation in the number of tasks one performs).

Figure 6 shows the graph density of groups with different team size. A clear trend can be observed that the graph density distribution of the real-world author-task bipartite graphs is more and more divergent from the density expected by the null model when team size grows. Specifically, the author-task graphs in real collaboration tend to be sparser and sparser than expected in the null model, which might suggest a stronger degree of labor division in larger teams.

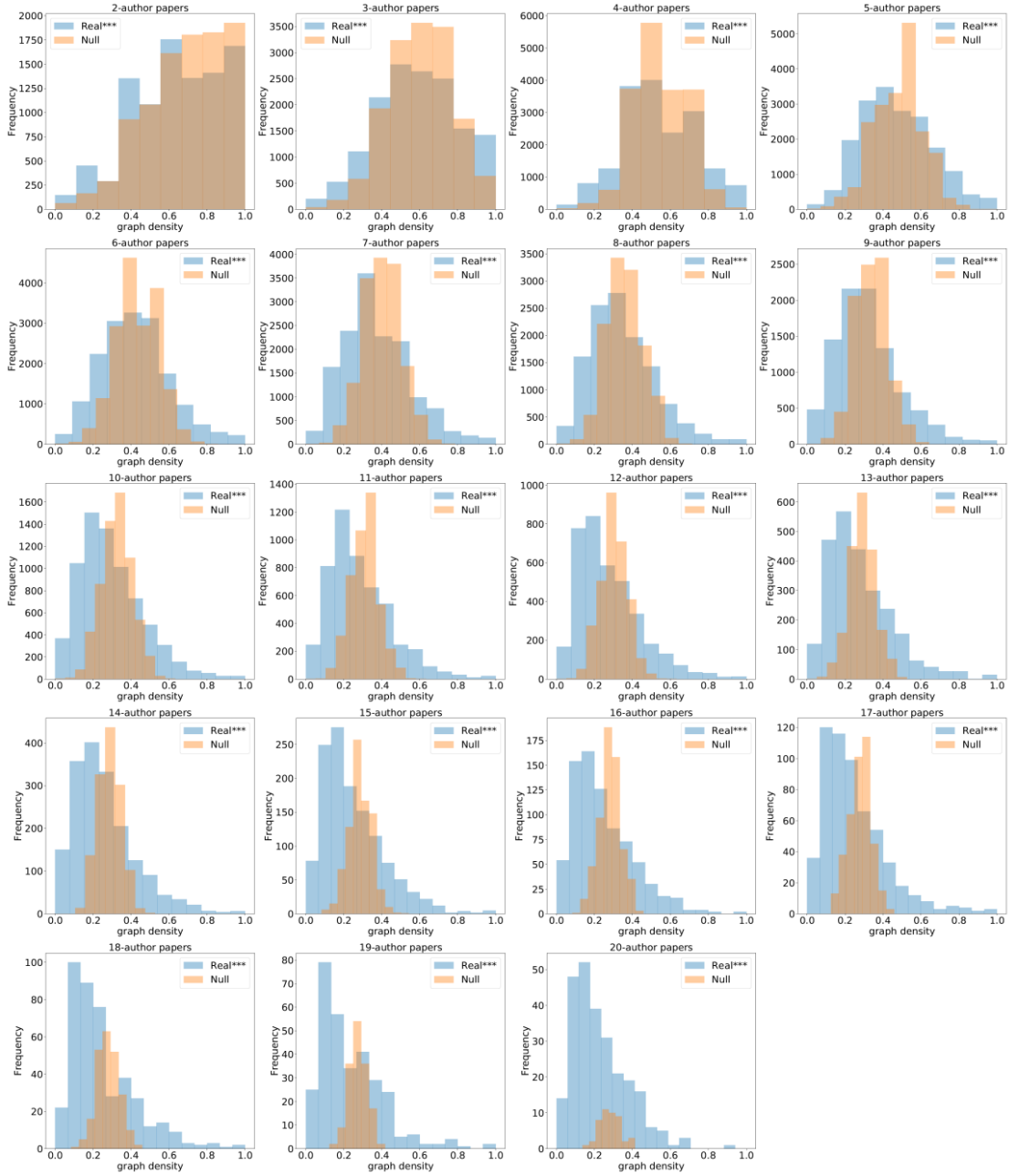

**Figure 6. Graph density against null model by team size (\* on the legend denotes the *p*-value from Kolmogorov–Smirnov test against the null model, \* for *p*-value <0.05, \*\* for *p*-value <0.01, \*\*\* for *p*-value <0.001).**

## *4.2. Quantifying types of collaborators*

In this section, we examine the prevalence of the three types of collaborators. First of all, we examine how many articles involve these three types of collaborators. We calculate the ratio of the articles containing *collaborator type $c_i$*, given *team size*

*k*, $PR_{C_i}^k$, using Formula (3) as follows:

$$PR_{C_i}^k = \frac{N_{C_i}^k}{N^k}, \quad c_i \in \{specialist, versatiles, team-player\}, \qquad (3)$$

where $N^k$ is the number of articles with *k* authors, and $N_{C_i}^k$ is the number of articles with *k* authors that contain *collaborator type $c_i$*.

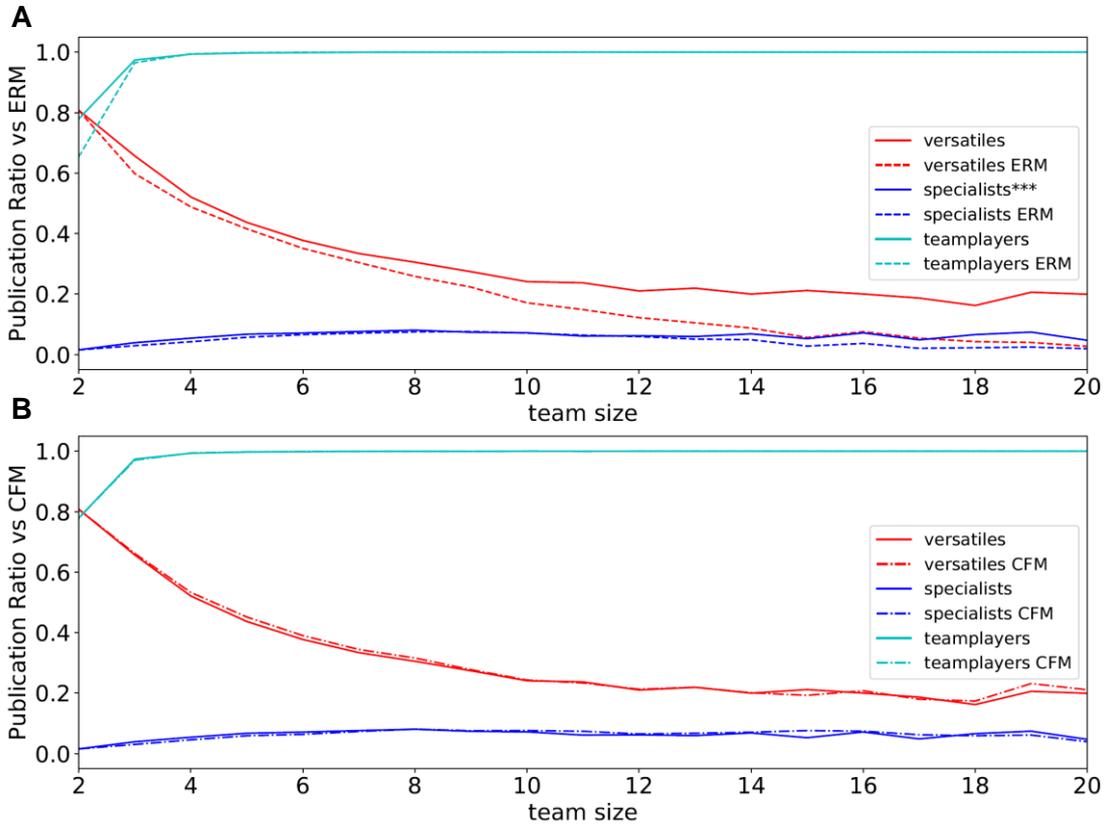

**Figure 7. Frequency Distribution of papers containing given collaborator type by team size: (A) against ERM null model; and (B) against CFM null model (\* on the legend denotes the *p*-value from K-S test against the null model, \* for *p*-value <0.05, \*\* for *p*-value <0.01, \*\*\* for *p*-value <0.001).**

Figure 7 shows that most articles have team-players, and that is expected by both null models (in A and B). There is a slight increase in the number of articles with specialists as team size increases. The articles with versatiles become less common as the size of teams increases, and the final ratio of such articles stabilizes around 25%.

When compared with the ERM null model, which shuffles author and task nodes in author-task bipartite graphs, it is found that non-team-players are more common in real-world collaborations than expected. More articles involve versatiles in real scientific collaborations. Specialists, instead of disappearing in larger teams as ERM suggests, keep playing a role in teams whose size varies from two to 20. It might suggest that non-team-players are associated with special and prevalent types of contributions in scientific collaborations.

Our result also suggests that the actual prevalence of each author type closely matches the expectation from the CFM null model, which shuffles authors' specific contributions in the bipartite graphs. It indicates that the degree sequence—how many tasks are performed by each author—accurately reproduces the co-contribution patterns that are observed.

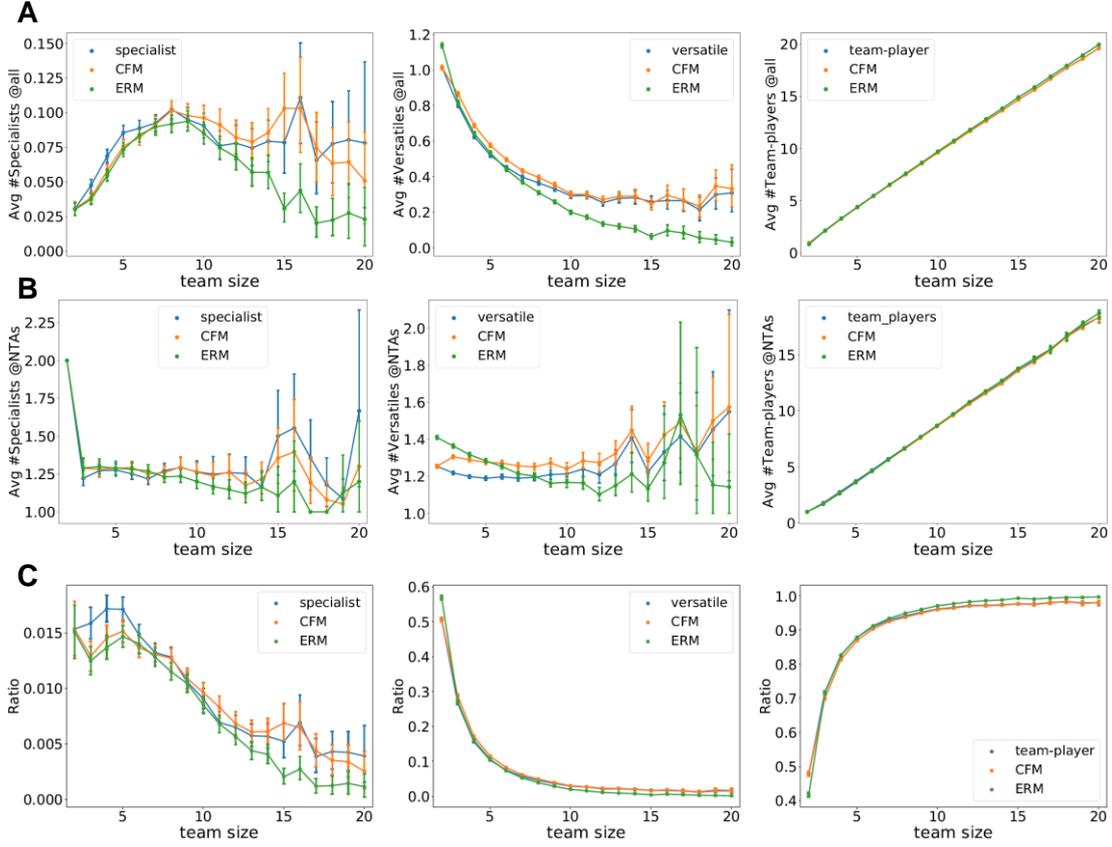

**Figure 8. (A)Types of collaborators in publications; (B) Team structure in NTAs (non-team-player-involved articles); and (C) Population ratio of types of collaborators. From left to right in each sub-plot are specialists, versatiles, and team-players; error bars in plots represent 95% confidence intervals generated via 10,000 bootstrapping iterations.**

To observe three *types of collaborators' existence in scientific collaborations*, we calculate the average number of each type of collaborators $C_i$ in teams by team size $k$, using Formula (4) as follows:

$$AC_i^k = \frac{TC_i^k}{N^k}, \quad c_i \in \{specialist, versatile, team-player\}, 2 \leq k \leq 20, \quad (4)$$

where $N^k$ is the number of teams with team size $k$, and $TC_i^k$ is the total number of *collaborator $i$* in *k*-authored publications. The results in Figure 7-A suggest that, on average, each paper contains around 0.075 specialists, 0.35 or more versatiles, and the rest team-players when the team size is greater than five. Specifically, when the team size is smaller than eight, the number of specialists increases along with the increase of team size and peaks at 0.1, which is well captured by both our null models. When team size continues growing, the real number of specialists fluctuates around the top, whereas that expected by ERM starts to diminish (shown in Figure 8-A-Left). Versatiles are more prevalent than specialists in scientific collaborations, especially among smaller teams. When the team grows larger, the average number of versatiles continues decreasing to 0.35 per paper then maintains stable. Besides that, despite the decreasing average number of versatiles in teams, it is still more than expected by ERM, in which the average number of versatiles keeps declining when the team size is larger than ten. Team-players dominate the participation in scientific collaborations, which is well captured by our two null models. The ERM null model only keeps the number of task assignments, while the CFM model sets some rules of labor division since it restricts how many tasks one author would participate and how many participants are involved for each individual task. By comparing the figures of the real situation in Figure 8-A with the two null models, we could see that there exist strong division of labor and that there are still more specialists and versatiles than we would expect from the random case among the collaborators. A downtrend of the number of versatiles along with the increase of team size is understandable, since the total number of tasks will not have an unlimited grow, so some authors may collaborate more with others when there are more team members. However, the slight increase or unchanged number of specialists demonstrates that there always exist some tasks that should be completed individually; the existence of specialists is important even in the environment of heavy collaboration. The distinction between the figures of non-team-players for the real-world collaborations and ERM indicates that the existence of non-team players is not because of small teams or limited labors in scientific collaborations but for particular purposes left for us to uncover. We are more interested in understanding *the structure of scientific teams* when they involve *heterogeneous collaborators*. So we exclude all the articles which were collaborated by only team-players and plot the average number of three different collaborator types among the remaining in Figure 8-B. In general, the team structure is quite stable with non-team-players among all different team sizes: 1.25 specialists on average, 1.3 versatiles on average, and with the rest team-players. More than often, a team includes one or two non-team-players to perform their research. Despite that, when teams grow larger than 15 participants, more non-team-players, specialists or versatiles, could contribute to the teams (suggested by the error bars). Team-players, similarly, are still dominating a team. The null models, however, do not show great disparity from the real collaborations in the team structures. It indicates when teams include non-team-players, there is no big variance among teams, especially for smaller teams whose size are less than ten.

We continue our focus on the overall *population of the three types of collaborators* among all the publications. We modify Formula (4) and calculate $RC_i^k$, the ratio of *collaborator $C_i$* given by *team size (k)*, using Formula (5) as follows:

$$RC_i^k = \frac{TC_i^k}{k \times N^k}, \quad c_i \in \{specialist, versatile, team-player\}, 2 \leq k \leq 20, \tag{5}$$

where $N^k$ is the number of teams with team size $k$, and $TC_i^k$ is the total number of *collaborator i* in *k*-authored publications. The results of Formula (4) is the results from Formula (3) normalized by team size accordingly (shown in Figure 8-C). Since the authors in our whole dataset are not disambiguated, the population character of these collaborators reflects how frequently a certain role (as three types of collaborators) have been played in scientific collaborations.

Figure 8-C demonstrates that non-team-players are the minority in scientific collaborations, as suggested above, especially for specialists. In particular, when team size grows, the ratio of specialists among collaborators drops from 15 percent to 0.5 percent then remains stable; the ratio of versatiles also falls from 55 percent to 3

percent and remains stable. Team-players, on the contrary, show an opposite trend, keeping increase from around 45 percent to around 95 percent. Both our two null models also roughly capture this trend.

To sum up, team-players are the major collaborators in scientific collaboration. Non-team-players are the minority, but they widely exist in small teams (a size no more than five) and also exist in larger teams (a size larger than five) with a relatively small and stable ratio.

A possible reason for the observations is that more team members enable division of labor and specialization (Smith, & McCulloch, 1838) rather than wiping out non-team-players. Some of the tasks performed by versatiles in smaller teams can be distributed to extra team members, accounting for more team-players. Regarding specialization, some team members can focus on particular tasks when more members are added to the team. For instance, in a dyatic team between advisor and advisee, besides supervision, the advisor may also need to take up some tasks such as writing and data analysis to accelerate the research progress. When more collaborators get involved, the advisor may spare more time and only focus on the supervision of advisees and funding application. Other collaborators can share the burden of the advisor (Bray et al., 1978) when the advisor could be a specialist and the other collaborators can function as team-players. Besides, specialized collaborators can be invited to the team to perform some special tasks as Specialists (or Specialists proposed in (Belbin, 2012)). The benefit of this evolution—division of labor and specialization—can increase the productivity of a team. On the other, however, more collaborators could bring the so-called "Ringelmann effect" (Ingham, Levinger, Graves, & Peckham, 1974) or "social loafing" (Earley, 1993), which means collaborators of a team tend to become increasingly less productive as the size of their team increases. However, this increasing tendency of specialization reaches saturation instead of excessively extending, which might be taken as the consideration of huge coordination cost that specialization may lead to (Becker & Murphy, 1992).

Randomly assigning tasks to authors (in ERM) leads to more non-team-players in smaller teams (≤10) and less in larger teams (≥15). By contrast, in real scientific collaborations, non-team-players maintain a relatively stable ratio in smaller teams and also exist in larger teams. Such existence is surprising in larger teams since adequate human resource facilitates us to perform tasks collaboratively to achieve seemingly efficiency and effectiveness.

Besides, it is worth noticing that versatiles tend to be more favorable than expected in scientific teams suggested by Figure 7. On the contrary, fewer versatiles in the ERM plot may imply that team-players are more welcome when they can also work independently as versatiles. The possible reason for this can be a moderate degree of specialization improves the efficiency of the collaboration when some tasks are performed alone and some collaboratively (Becker & Murphy, 1992).

### 4.3. Understanding Collaborators' Tasks

After quantitatively describing the prevalence of the three types of collaborators in our dataset, here we analyze their characteristics by examining the tasks in which they participated. We look at the most common five tasks (e.g., Corrêa et al., 2017; Larivière et al., 2016) as well as the other less frequent tasks. Using the data generated by our null models used in the previous section, i.e., CFM and ERM, we can also investigate the different patterns in task distributions between real collaborations and two random scenarios for different purposes. CFM controls authors' and tasks' degree sequence in an author-task network; thus, the differences from real collaborations highlight the differences in task-performing, which will suggest prevailing patterns of different types of authors in reality. ERM only controls the number of edges in the networks. The corresponding results can be used to exam whether these task patterns can be generated randomly. We extract the top 100 most frequent tasks for each type of collaborators from the three data sets (including the two generated null model data sets) and consolidate similar tasks. As a result, 52 unique tasks are obtained.

## Five Common Tasks

Figure 9 presents the five most common tasks in the author contribution statement from *PLoS*. The radar plots suggest that although the three types of authors all engage in the five most common tasks, the emphasis varies. For instance, Specialists "contribute reagents/materials/analysis tools" much more often than expected while "performed experiments" much less than expected (Figure 9-A). This result suggests that "reagents/materials/analysis tools" task can be more easily isolated to a single person than other tasks, and that it is rare for an author to just perform experiments and not participating in other tasks with others, indicating the central role of the task of performing experiments in scientific studies. Compared with null models, specialists make much less contribution to performing experiments. It might indicate that specialists are not usually implementers but toolmakers in a team. On the contrary, versatiles and team-players show more balanced contributions to all the five common tasks. Despite that, the null models suggest that versatiles contributed much less to the five common tasks than expected, indicating their emphasis on less common tasks. Team-players show indifferences in null models, which could be attributed to their massive population in our data set.

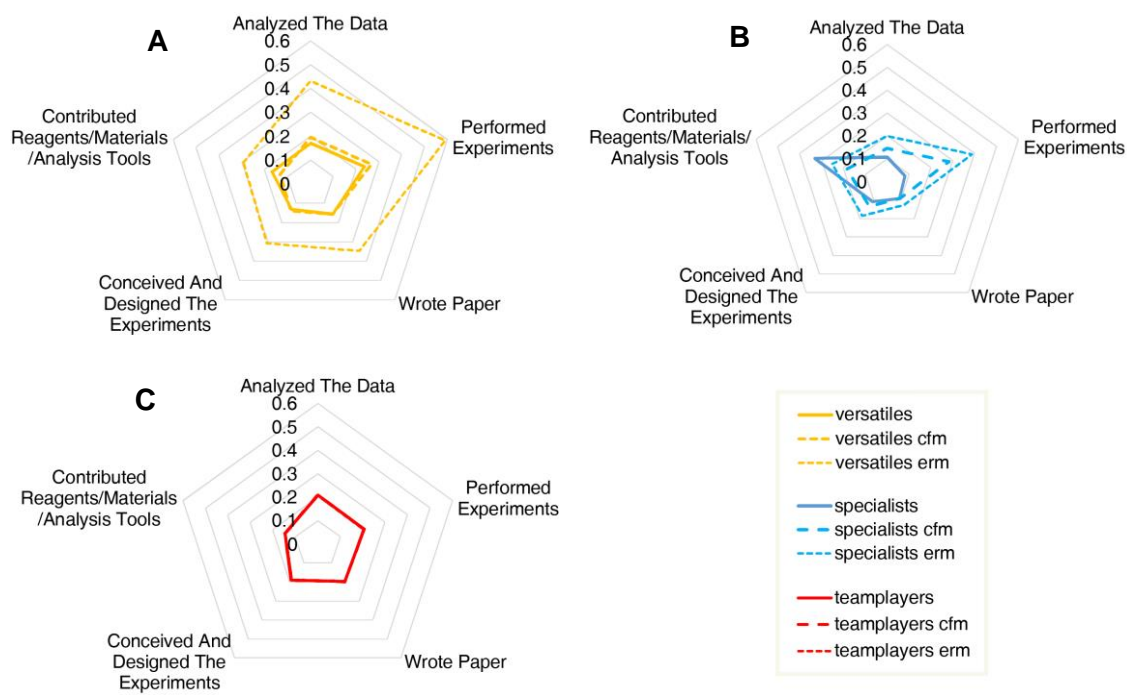

**Figure 9. Common tasks performed by the three types of collaborators: A, Specialists; B, Versatiles; C, Team-players.**

## Less frequent tasks

Figure 10 and Figure A1 (in Appendix) show the participation patterns of the less frequent tasks. Greater disparities emerge, which have been overlooked by existing studies that focus only on a few core tasks (e.g., Corrêa et al., 2017; Larivière et al., 2016). First, team-players participate in these activities much less frequently, except for "approve the paper", which usually occurs at the final stage of their research. "Data interpretation" is another task that team-players do more frequently than non-team-players. The possible reason for this is that data interpretation is interdependent with data analysis, which is team-players' major task in Figure 9-C. Similar to Figure 9-C, both null models capture almost identical patterns for less frequent tasks.

Second, Specialists show a strong tendency to take the tasks like "review paper", "revise paper", and "supervised the research." This might indicate that specialists can be senior investigators in teams. Some of the following tasks, like "principal investigator" and "provided guidance", also suggest our inference. The CFM confirms that specialists contributed more to these tasks as senior authors than expected, such as "revise paper" and "supervise the research". In addition, Specialists also perform tasks that may not be that crucial, such as "collected data" and "collected samples." This type of tasks may also suggest that specialists can be mild collaborators (Hara et al., 2003). As suggested by the following tasks as well, they also take charge of "database management" and "provided technical support" (In Figure A1). CFM also confirms this tendency. Versatiles tend to partake in authority-intensive and idea-intensive tasks. For example, most of the funding related tasks are versatiles' work. Designing software and designing models are usually versatiles' tasks. It is confirmed by our CFM. We may infer that some versatiles are either leaders of certain projects or chief authors of the studies.

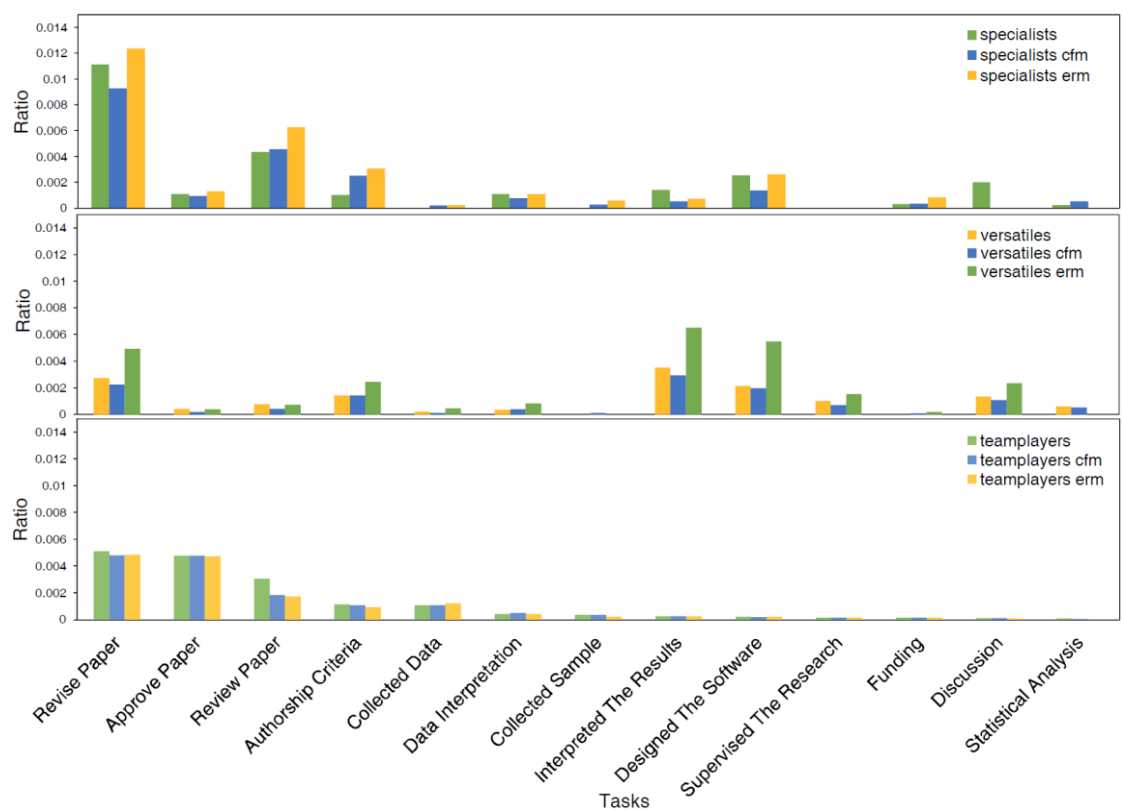

**Figure 10. Less frequent tasks performed by the three types of collaborators (More information about less frequent tasks is presented in Figure A1).**

# 5. CONCLUSION

In this study, we proposed a refined approach—author contribution network for each publication based on the author contribution statements embedded in the body of manuscripts. It aims at better understanding scientific collaboration at the task assignment level. More than 130 thousand articles were collected to perform our analyses.

The results suggested that scientific collaboration within the team could be diverse. Inspired by the concepts of division of labor and specialization by Adam Smith, we identified three types of collaborators in the author co-contributorship network: they are called specialists, versatiles, and team-players. The three types of collaborators form diverse teams and contribute to publications in various ways.

Team-players are the backbones in scientific studies. They usually contribute to the five common tasks (data analysis, performing experiments, writing papers, and contribute materials and tools). They seldom take up tasks with authorities (i.e., providing funding or project supervision). Versatiles are not that common in a team as team-players are. They are usually those who connect collaborators in a team (with edges to other collaborators) and do well in all five common tasks with a specialty in performing experiments. They are also featured with a high level of authorities in teams. For example, study supervision and obtaining funding are their dominant tasks among the less frequent ones. Specialists are special since they usually maintain such a small population across different team sizes. Larger teams cannot eliminate them. Besides, they put themselves in a distinct position of performing collaborations. They are usually those who contribute tools, materials, and special supports. These supports can either sign their authority in a team, like providing financial support, or their blur figures, such as technical support.

These observations can help in various ways in the future: author credit assessment, team structure optimization, and candidate projects assessment guidance. Usually, author credit is given by the authors' byline position either evenly or differently (Stallings et al., 2013). These operations can be problematic sometimes when the collaboration between authors are not well assigned (e.g., Sauermann, & Haeussler, 2017). Given the contributions the authors performed, their roles and credits could be given more fairly with a well-defined system of contribution scoring.

Teams vary but only a few of them succeed. And they are not simple combinations of the geniuses but of diverse roles and complementary skills (Amabile et al., 2001; Belbin, 2012). Our work might signify a way to build a scientific team with consideration of members' most frequent tasks in their earlier studies given limited resources and expense.

Similarly, the co-contributorship network may also help us to find patterns of success based on the characteristic of the team members, task division and assignment, and specialization within the teams. Thus, the funding agencies can achieve better assessment with the team structure of the proposals according to their publication history and their roles in these studies.

However, with so many aforementioned potentials, this study has several limitations. One is that this study so far takes each contribution equally while the criteria could vary across disciplines. How to weigh different types of contributions based on specific applications could become an interesting area to study. Second, the dataset of this study mainly comes from natural science, especially biology. But , among the five common tasks "contributed reagent/materials/analysis tools" is not common in social science.. Extending data to social science or other fields is an important next step to follow. Third, this study has only studied a collaborative team associated with one single article. It does not investigate the perspective that a researcher joining different teams and playing various roles. After author names have been disambiguated, we can address fascinating questions, such as, . how a scholar's career evolves based on his/her team roles in collaborations. Fourth, Collaboration is becoming international. Taking nationalities, culture barriers, institutional prestige, skill diversity into the consideration, the division of labor can be further expanded to the social, behavioral, and political arena which makes it complicated yet exciting to pursue.

## Acknowledgments

The authors would thank the anonymous reviewers for their constructive suggestions for improving this manuscript.

# Appendix

**Figure A1. less frequent tasks performed by different collaborators**

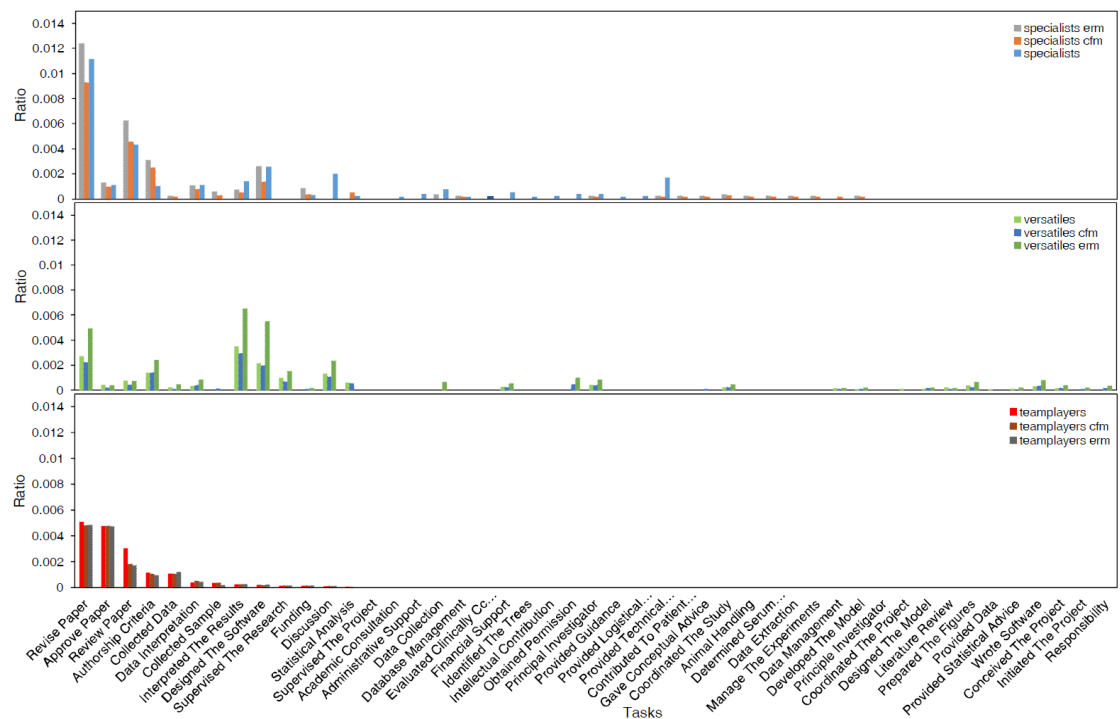